\documentclass[preprint]{aastex}
\newcommand{\iint}{\int\!\!\!\!\int}
\newcommand{\bm}[1]{\mbox{\boldmath$#1$}}
\begin{document}
\title{
Continuous Limit of Multiple
Gravitational Lens Effect and Average Magnification Factor}
  \author{Hiroshi Yoshida\altaffilmark{1}, Kouji
   Nakamura\altaffilmark{2} and Minoru Omote\altaffilmark{3}}
   \altaffiltext{1}{Fukushima Medical University, Fukushima-City 960-1295, 
   Japan; E-mail:\texttt{yoshidah@fmu.ac.jp}}
   \altaffiltext{2}{National Astronomical Observatory, Mitaka 181-8588,
   Japan; E-mail:\texttt{kouchan@th.nao.ac.jp}} 
   \altaffiltext{3}{Keio University, Yokohama  223-8521, Japan;
   E-mail:\texttt{omote@phys-h.keio.ac.jp}}
\begin{abstract}
 We show that the gravitational magnification factor averaged over all 
 configurations of lenses in a locally inhomogeneous universe satisfy a 
 second order differential equation with redshift $z$ by taking the 
 continuous limit of multi-plane gravitational lens equation (the number 
 $N_\mathrm{L}$ of lenses $\to\infty$) and that the  gravitationally magnified 
 Dyer-Roeder distance in a clumpy universe  becomes to that of the 
 Friedmann-Lema\^{\i}tre universe for arbitrary values of the density 
 parameter $\Omega_{0}$ and of a mass fraction $\bar{\alpha}$ 
 (smoothness  parameter). 
\end{abstract}
\section{Introduction}
A light ray propagation in a locally inhomogeneous universe has been 
investigated by many authors using analytical and/or numerical methods
(e.g., \citealt{OY90,YO92};\citealt*[][and references therein]{SEF92}). 
By taking gravitational lens effects into account,  
\cite{Wein76} showed that in a case of the low deacceleration parameter 
$q_0=\Omega_0/2-\lambda_0$ ($\Omega_0, \lambda_0$ are the  
density parameter and the cosmological constant, respectively) an
average flux from sources in a clumpy universe  is equal to the flux in 
the Friedmann-Lema\^{\i}tre universe (flux conservation). 
For a more general value of $\Omega_0$, some authors 
\citep[e.g.,][]{ES86,Pea86} discussed the 
gravitational magnification probability function by assuming the 
flux conservation.
\par
Recent observations on high-redshift Type Ia 
supernovae \citep{SNIa} and on the cosmic microwave background 
\citep[CMB,][]{WMAP} suggest  that our universe is 
accelerating in expansion rate. In such a situation, the 
deacceleration parameter may be no longer 
small. Then it is needed to consider the light ray propagation in a more 
general inhomogeneous universe model with arbitrary values of $\Omega_0$ 
and $\lambda_0$ for sources with high-redshifts. 
\par
In a  case with an arbitrary $\Omega_0$, 
we have to take magnification effects by multiple lenses into account. 
In studies of this problem the multi-plane lens theory has been used 
both in  analytical approximations 
\citep{Pea86,Isaac89,SW88a,Wu90,Marchan,SS94} and in  numerical 
simulations (\citealt{Refs70,SW88b,WT90,Rauch90};~\citealt*{Lee97,PMMF01}). 
In analytical studies many authors \citep{VO83,Pei93,Schneider93} have 
assumed that the total magnification by lenses can be approximately 
given by a product of the magnifications of individual lens, and have 
obtained statistically the total magnification by gravitational lenses 
distributed at random in the universe. 
\par
In this paper we consider the total gravitational magnification factor 
averaged with all configurations of lenses distributed at random in a 
locally inhomogeneous universe and discuss the continuous limit (the 
number $N_\mathrm{L}$ of lens planes $\to\infty$) in which lens planes approach to 
be continuously distributed. In \S 2 the multi-plane lens theory is 
briefly reviewed and an average magnification matrix is obtained in \S 
3. In \S 4 the continuous limit of the magnification matrix is considered. 
We show that in this limit the average magnification factor satisfies 
a second order differential equation and that a angular diameter distance 
multiplied by $\left<\mu(z)\right>^{-1/2}$ ($\left<\mu(z)\right>$: 
average magnification factor) reduces to the angular diameter distance 
of the homogeneous universe (the Friedmann--Lema\^{\i}tre universe).  
\section{Multi-plane lens equation}
\begin{figure}[tb]
\plotone{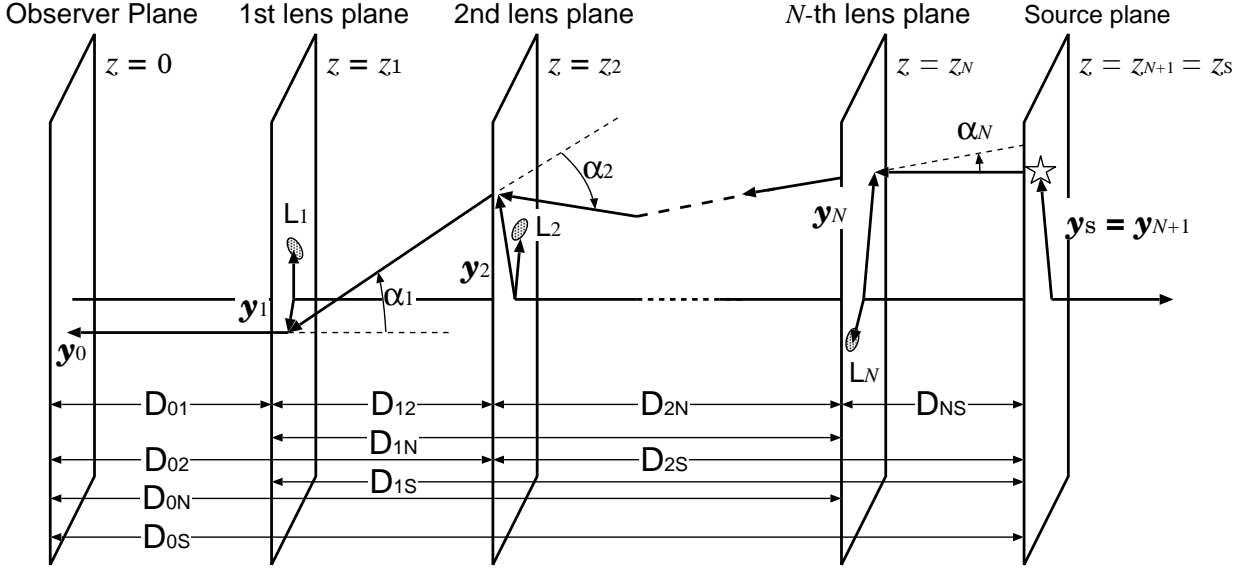}
\caption{Geometry of multi-plane lens system: $\mathrm{L}_i$ is the $i$-th 
lens. Each lens plane is perpendicular to observer's line of sight. The 
origin of each lens plane is set on the line of sight. The light ray 
observed at $\bm{y}_0$ in the observer plane crosses at $\bm{y}_i$ in 
the $i$-th lens plane. $D_{ij}$ denotes the angular diameter distance 
from the $i$-th lens  plane to the $j$-th lens plane (the observer is in the 
the $0$-th lens plane and the source is in the ($N+1$)-th lens 
plane).\label{fig:mlp}}  
\end{figure}

We will give a brief review of the multi-plane lens equation in this section.
Suppose that $N_\mathrm{L}$ lenses are randomly distributed at redshifts 
$z_i(0\le z_{1}<z_{2} \cdots <z_{N_\mathrm{L}})$, and that 
$z_\mathrm{S}=z_{N_\mathrm{L}+1}>z_{N_\mathrm{L}}$ is a redshift of the 
source (see Fig. \ref{fig:mlp}). 
The multi-plane lens equation for the source is given by
\begin{equation}
\bm{y}_\mathrm{S}={D(0;z_\mathrm{S})\over 
D(0;z_1)}\bm{y}_1-\sum_{i=1}^{N_\mathrm{L}}D(z_i;z_\mathrm{S})
\bm{\alpha}_i(\bm{y}_i) ,\label{eq:mlpSEF}
\end{equation}
where $\bm{y}_\mathrm{S}$ denotes the position 
vector of the source at the source plane and 
$\bm{y}_i$ is the position vector of the light ray at the $i$-th lens plane
\citep{SEF92}.
In equation (\ref{eq:mlpSEF}) $D(z_i;z_j)$ is the angular diameter 
distance from  the $i$-th lens plane at $z_i$ to the $j$-th lens plane at $z_j$. 
The deflection angle $\bm{\alpha}_i(\bm{y}_i)$ at the $i$-th lens plane is 
given by 
\begin{equation}
\bm{\alpha}_i(\bm{y}_i)=
{4G\over c^2}\iint_{{\cal S}_i}d^2\bm{y}_i'\Sigma_i(\bm{y}_i')
{\bm{y}_i-\bm{y}_i'\over|\bm{y}_i-\bm{y}'_i|^2},
\label{eq:alpha}
\end{equation}
where $\Sigma_i(\bm{y}_i')$ is a surface mass density of the $i$-th lens 
and ${\cal S}_i$ denotes the observed region on the $i$-th lens plane.
\par
We should notice that an image at $\bm{y}_i$ on the $i$-th lens plane 
could be regarded as a ``source'' by the foreground lenses. Therefore 
the multi-plane lens equation for the ``source'' at $\bm{y}_i$ can be 
rewritten as follows: 
\begin{equation}
\bm{y}_i=
{D(0;z_i)\over D(0;z_1)}\bm{y}_1-\sum_{j=1}^iD(z_j;z_i)\bm{\alpha}_j(\bm{y}_j).
\label{eq:mlpi}
\end{equation}
In the following we use new variables $\bm{\theta}_i=\bm{y}_i/D(0;z_i)\ 
(i=1,\cdots,N_\mathrm{L}+1=\mathrm{S})$ which denote 
the angular position of the light ray in the $i$-th lens plane. Using a 
dimensionless angular diameter distance $d(z_i;z_j)=D(z_i;z_j)/(c/H_0)$, 
we can rewrite equation (\ref{eq:mlpi}) as follows:
\begin{equation}
\bm{\theta}_i=\bm{\theta}_1-\sum_{j=1}^i{d(z_j;z_i)\over d(0;z_i)}
\bm{\alpha}_j[D(0;z_j)\bm{\theta}_j].
\label{eq:mlptheta}
\end{equation}
\par
Using the $\chi$-function introduced by \citeauthor{SEF92} (see equation 
[\ref{eq:chijfunc}] in Appendix A), the distance $d(z_j;z_i)$ from the 
$j$-th lens plane to the $i$-th lens plane is given by 
$d(z_j;z_i)=(1+z_j)d(0;z_j)d(0;z_i)(\chi_j-\chi_i)$, then equation 
(\ref{eq:mlptheta}) is rewritten in the following expression
\begin{equation}
\bm{\theta}_i=
\bm{\theta}_{i-1}-(\chi_{i-1}-\chi_i)
\sum_{j=1}^{i-1}(1+z_j)\tilde{\bm{\alpha}}_j(\bm{\theta}_j),
\label{eq:mlpii_1}
\end{equation}
where 
\begin{equation}
\tilde{\bm{\alpha}}_j(\bm{\theta}_j)
=d(0;z_j)\bm{\alpha}_j[D(0;z_j)\bm{\theta}_j]
={4G\over cH_0}d^2(0;z_j)
\iint_{{\cal D}} d^2\bm{\theta}_j'\Sigma_j[D(0;z_j)\bm{\theta}_j']
{\partial\ \over\partial\bm{\theta}_j}\ln|\bm{\theta}_j-\bm{\theta}_j'|,
\label{eq:Alpha_i}
\end{equation}
and $\bm{\theta}'_j$ and ${\cal D}$ denote the angular coordinate on the 
$j$-th lens plane and the observed region, respectively. An expression 
similar to equation (\ref{eq:mlpii_1}) has been given by \cite*{PLW01}. 
This recurrence formula (\ref{eq:mlpii_1}) determines iteratively the 
``source'' position $\bm{\theta}_i$ in terms of $\bm{\theta}_j (j<i)$ and is 
useful in a numerical experiment based on the ray tracing method. 
An equivalent equation to equation (\ref{eq:mlpSEF})  
can be also obtained from the Fermat principle \citep{BL86,Kovn87}:
\begin{equation}
\chi_{i,i+1}\left(\bm{\theta}_{i+1}-\bm{\theta}_i\right)
=\chi_{i-1,i}\left(\bm{\theta}_i-\bm{\theta}_{i-1}\right)
-(1+z_i)\tilde{\bm{\alpha}}_i(\bm{\theta}_i),
\label{eq:mlpBN}
\end{equation}
where $\chi_{i,j}=[\chi_i-\chi_j]^{-1}$.
\par
Now we give the magnification matrix $A_{\mathrm{S},N_\mathrm{L}}$ 
in the case of the multi-plane lensing by using equation (\ref{eq:mlpSEF}) and
(\ref{eq:mlptheta}) as
\begin{equation}
A_{\mathrm{S},N_\mathrm{L}}\equiv
{\partial \bm{\theta}_\mathrm{S}\over\partial \bm{\theta}_1}
=I-\sum_{i=1}^{N_\mathrm{L}}{(1+z_i)\over\chi_{i,\mathrm{S}}}\tilde{U_i}A_i,
\label{eq:MagA}
\end{equation}
where $I$ is the $2\times2$ unit matrix and $\tilde{U}_i$ and $A_i$ are 
matrices defined by
\begin{equation}
\tilde{U}_i\equiv{\partial \tilde{\bm{\alpha}}_i\over \partial \bm{\theta}_i},\quad 
A_i\equiv{\partial \bm{\theta}_i\over\partial\bm{\theta}_1}.
\label{eq:UA}
\end{equation}
We should notice that $\tilde{U}_i$ defined in 
terms of $\tilde{\bm{\alpha}}_i$ is slightly different from the matrix 
$U_i$ $(\equiv\partial\bm{\alpha}_i/\partial\bm{\theta}_i)$ in \cite{SEF92}. 
While $U_i$ in their  definition depends on the redshift $z_\mathrm{S}$ 
of the source, $\tilde{U}_i$ in our definition does not since 
$\tilde{\bm{\alpha}}_i$ is independent of $z_\mathrm{S}$.
By virtue of equations (\ref{eq:mlptheta}), (\ref{eq:MagA}) and (\ref{eq:UA}), 
we obtain the following form:
\begin{equation}
A_{\mathrm{S},N_\mathrm{L}}=I+\sum_{k=1}^{N_\mathrm{L}}(-)^k\sum_{i_1=1}^{i_0}
\sum_{i_2=1}^{i_1}\cdots\sum_{i_k=1}^{i_{k-1}}{(1+z_{i_1})(1+z_{i_2})\cdots 
(1+z_{i_k})\over\chi_{i_k,i_{k-1}}\cdots\chi_{i_2,i_1}\chi_{i_1,i_0}} 
\tilde{U}_{i_1}\tilde{U}_{i_2}\cdots\tilde{U}_{i_k},
\label{eq:MagAS}
\end{equation}
where $i_0 ~(=N_\mathrm{L}+1)> i_1> i_2>\cdots> i_{k-1}> i_k\ge1$,
and the matrix $\tilde{U}_i$ can be expressed as
\begin{eqnarray}
\tilde{U}_i(\bm{\theta}_i)&=&{4G\over cH_0}d^2(0;z_i)\iint_{{\cal D}}
d^2\bm{\theta}'\Sigma_i[D(0;z_i)\bm{\theta}']\tilde{U}_i'(\bm{\theta}_i-\bm{\theta}'),
\label{eq:Ui}\\
\tilde{U}_i'(\bm{\eta})&=&\left(\begin{array}{cc}
\pi\delta^2(\bm{\eta})-\Gamma_1(\bm{\eta})&-\Gamma_2(\bm{\eta})\\
-\Gamma_2(\bm{\eta})&\pi\delta^2(\bm{\eta})+\Gamma_1(\bm{\eta})\\
\end{array}
\right),
\end{eqnarray}
and 
\begin{equation}
\Gamma_1(\bm{\eta})={\eta_x^2-\eta_y^2\over|\bm{\eta}|^4},\quad 
\Gamma_2(\bm{\eta})={2\eta_x\eta_y\over|\bm{\eta}|^4}.
\end{equation}
\par
Recurrence formulae of the magnification matrix given by
equation (\ref{eq:mlpii_1}) or (\ref{eq:mlpBN}) are also written as:
\begin{eqnarray}
&&A_i=A_{i-1}-\frac{1}{\chi_{i-1,i}}\sum_{j=1}^{i-1}(1+z_j)\tilde{U}_jA_j,
\label{eq:Magii_1}\\
&&\chi_{i,i+1}\left(A_{i+1}-A_i\right)=\chi_{i-1,i}\left(A_i-A_{i-1}\right)
-(1+z_i)\tilde{U}_iA_i.\label{eq:MagBN}
\end{eqnarray}
\section{Average Magnification Matrix}
In this section the universe is assumed to be a locally inhomogeneous, 
on-average homogeneous and isotropic universe in which a mass fraction 
$\bar{\alpha}$ (smoothness parameter) of the mean matter density $\bar{\rho}(z)$
is smoothly distributed, while a fraction $(1-\bar{\alpha})\bar{\rho}(z)$ is 
concentrated into clumps distributed at random. The angular diameter 
distance $D(z;z')$ of this universe from a redshift $z$ to another 
redshift  $z'$ satisfies the Dyer--Roeder equation (\ref{eq:GD}) with 
$0\le\bar{\alpha}<1$ \citep{DR73}.
\par
In this universe a light ray passes through the space with the smoothly 
distributed mass density $\bar{\alpha} \bar{\rho}(z)$ and is 
gravitationally affected several times by clumps (lenses) located near 
the light path, in general. 
Since the gravitational magnification factor for the light ray depends on 
the distribution of lenses near the light path, we cannot 
discuss the individual gravitational magnification factor of a source 
without the knowledge about the configuration of lenses near 
the light ray from the source.
Nevertheless it is meaningful to estimate an average 
gravitational magnification factor for light rays which travel in 
various regions of the inhomogeneous universe, because  the 
factor plays an important role in the 
theoretical analysis of observed data such as $m-z$ relation.
\par
In the following we consider only the gravitational 
magnification factor 
$\left<\mu\right>=\det\left<A_{\mathrm{S},N_\mathrm{L}}\right>^{-1}$ 
averaged over all distributions 
$\{\bm{\xi}_1,\cdots,\bm{\xi}_{N_\mathrm{L}}\}$ of lenses on each lens 
plane ($\bm{\xi}_i$: center of the $i$-th lens) in the locally 
inhomogeneous universe defined by
\begin{eqnarray}
\left<A_{\mathrm{S},N_\mathrm{L}} \right>&\equiv&
{\iint_{\cal D}d^2\bm{\xi}_1\cdots\iint_{\cal D}}
d^2\bm{\xi}_{N_\mathrm{L}} A_{\mathrm{S},N_\mathrm{L}}(\bm{\xi}_1,\cdots,\bm{\xi}_{N_\mathrm{L}})
\biggl/{\iint_{\cal D}d^2\bm{\xi}_1\cdots\iint_{\cal D}}
d^2\bm{\xi}_{N_\mathrm{L}}\nonumber\\
&=&Q^{-N_\mathrm{L}}\iint_{\cal D}d^2\bm{\xi}_1\cdots\iint_{\cal D}
d^2\bm{\xi}_{N_\mathrm{L}} A_{\mathrm{S},N_\mathrm{L}}(\bm{\xi}_1,\cdots,\bm{\xi}_{N_\mathrm{L}}),
\label{eq:mASNL}
\end{eqnarray}
where $Q$ is the solid angle of 
the observed region $\cal D$. Here it should be noticed that to take all 
configurations of lenses into account means to consider observed regions 
in various directions as well as various lens distributions on the 
individual lens plane.
By virtue of equation (\ref{eq:MagAS}) the 
average magnification matrix is given by 
\begin{eqnarray}
\left<A_{\mathrm{S},N_\mathrm{L}}\right>
&=&I+\sum_{k=1}^{N_\mathrm{L}}
(-)^k\sum_{i_1=1}^{i_0}\sum_{i_2=1}^{i_1}
\cdots
\sum_{i_k=1}^{i_{k-1}}
{(1+z_{i_k})\cdots(1+z_{i_2})(1+z_{i_1})
      \over
\chi_{i_k,i_{k-1}}\cdots\chi_{i_2,i_1}\chi_{i_1,i_0}
 }\nonumber\\
&&\hspace{20mm}\times\left<\tilde{U}_{i_1}(\bm{\xi}_{i_1};\bm{\theta}_{i_1})\tilde{U}_{i_2}
(\bm{\xi}_{i_2};\bm{\theta}_{i_2})\cdots\tilde{U}_{i_k}(\bm{\xi}_{i_k};
\bm{\theta}_{i_k})\right>,
\label{eq:MASNL}
\end{eqnarray}
where $\tilde{U}_i(\bm{\xi}_i;\bm{\theta}_i)$ denotes the matrix 
for the $i$-th lens centered on $\bm{\xi}_i$ given by equation (\ref{eq:Ui}).
\par
As shown in Appendix B, if the lenses are distributed at random in the 
universe, i.e., if they does not correlate each other, the 
average of the product of matrices $\tilde{U}_i$ reduces to the 
product of the average matrices $\left<\tilde{U}_i\right>$, i.e.,
\begin{equation}
\left<\tilde{U}_{i_1}(\bm{\xi}_{i_1};\bm{\theta}_{i_1})
\tilde{U}_{i_2}(\bm{\xi}_{i_2};\bm{\theta}_{i_2})\cdots 
\tilde{U}_{i_k}(\bm{\xi}_{i_k};\bm{\theta}_{i_k})\right>
=\left<\tilde{U}_{i_1}(\bm{\xi}_{i_1};\bm{\theta}_{i_1})\right>
\left<\tilde{U}_{i_2}(\bm{\xi}_{i_2};\bm{\theta}_{i_2})
\right>\cdots\left<\tilde{U}_{i_k}(\bm{\xi}_{i_k};\bm{\theta}_{i_k})\right>. 
\label{eq:UUU}
\end{equation}
\par
We shall put $\bm{\theta}_i'\equiv\bm{\xi}_i+\tilde{\bm{\theta}_i}$ in 
equation (\ref{eq:Ui}) and assume that all lenses have the same mass 
profile, then all lenses have the same surface density 
$\tilde{\Sigma}(\tilde{\bm{\theta}}_i;z_i)=\Sigma[D(0;z_i)\bm{\theta}_i']$ 
which does not depend on the center $\bm{\xi}_i$ of the $i$-th lens. 
Under this assumption, equation (\ref{eq:Ui}) becomes to
\begin{equation}
\tilde{U}_i(\bm{\xi}_i;\bm{\theta}_i)={4G\over cH_0}d^2(0;z_i)\iint_{{\cal D}}
d^2\tilde{\bm{\theta}}_i\tilde{\Sigma}(\tilde{\bm{\theta}}_i;z_i)
\tilde{U}_i'(\bm{\theta}_i-\bm{\xi}_i-\tilde{\bm{\theta}}_i). 
\label{eq:Uip}
\end{equation}
and then the average matrix $\left<\tilde{U}_i\right>$ can be written by 
\begin{equation}
\left<\tilde{U}_i(\bm{\xi}_i;\bm{\theta}_i)\right>={G\over cH_0Q}d^2(0;z_i)
\iint_{{\cal D}}d^2\tilde{\bm{\theta}}_i\tilde{\Sigma}(\tilde{\bm{\theta}}_i;z_i)
\iint_{{\cal D}} d^{2}\bm{\xi}_i \tilde{U}_{i}^{\prime}
(\bm{\theta}_i-\bm{\xi}_i-\tilde{\bm{\theta}}_i).
\label{eq:UM}
\end{equation}
When the matrix 
$\tilde{U}_i^{\prime}(\bm{\theta}_i-\bm{\xi}_i-\tilde{\bm{\theta}}_i)$ is 
integrated with $\bm{\xi}_i$ in a large region, we have 
$\left<\tilde{U}'_i(\bm{\xi}_i;\bm{\theta}_i)\right>_{ab}=\pi \delta_{ab}/Q$ 
($a,b=1$ or $2$) since the shear terms $\Gamma_1$ and $\Gamma_2$ in 
$\tilde{U}_{i}^{\prime}$ vanish because of symmetry. Then we find  
\begin{equation}
\left<\tilde{U}_i(\bm{\xi}_i;\bm{\theta}_i)\right>={4\pi G\over cH_0Q}d^2(0;z_i)\iint_{{\cal D}}
d^2\tilde{\bm{\theta}}_i\tilde{\Sigma}(\tilde{\bm{\theta}}_i;z_i)I. 
\label{eq:MUi}
\end{equation}
\par
In equation (\ref{eq:MUi}) $\tilde{\Sigma}(\tilde{\bm{\theta}}_i;z_i)$ 
can be expressed in terms of the matter density 
$\rho[D(0;z_i)\bm{\theta}_i',Z_i]$ in the universe,   
where $Z_i$ is a coordinate along the line of sight
given by the cosmological time $T(z_i)$ and its present 
value $T(0)$ as $c[T(0)-T(z_i)]$. 
Since the smoothly distributed matter does not contribute to the 
deflection angle $\tilde{\bm{\alpha}}$ in equation (\ref{eq:Alpha_i}), 
we can find that the contribution to the magnification matrix comes from 
the inhomogeneous part of $\rho[D(0;z_i)\bm{\theta}_i',Z_i]$. Then the 
surface mass density of the $i$-th lens plane are expressed as 
\begin{equation}
 \Sigma[D(0;z_i)\bm{\theta}_i']=\delta_{\bar{\alpha}}\rho[D(0;z_i)
\bm{\theta}_i',Z_i]\cdot\bigl|c\Delta T_i\bigr|
={c\over H_0}\delta_{\bar{\alpha}}\rho[D(0;z_i)\bm{\theta}_i',Z_i]
{\Delta z_i\over (1+z_i)Y(z_i)}, \label{eq:Simga_rho}
\end{equation}
where 
\begin{equation}
 \delta_{\bar{\alpha}}\rho[D(0;z_i)\bm{\theta}_i',Z_i]
\equiv\rho[D(0;z_i)\bm{\theta}_i',Z_i]-\bar{\alpha}\bar{\rho}(z_i).
\label{eq:delta_rho}
\end{equation}
Then we find the mass on the $i$-th lens plane is given by
\begin{equation}
\iint_{{\cal D}}d^2\tilde{\bm{\theta}}_i\tilde{\Sigma}(\tilde{\bm{\theta}}_i;z_i)=Q(1-\bar{\alpha})\bar{\rho}(z_i)\frac{c\Delta z_i}{H_0(1+z_i)Y(z_i)}.
\label{eq:avMass}
\end{equation}
Using $\bar{\rho}(z)=(1+z)^{3}\bar{\rho}_0$ and 
$\bar{\rho}_0=3H_{0}^{2}\Omega_0/8\pi G$, equation (\ref{eq:MUi}) can be 
rewritten as 
\begin{equation}
\left<\tilde{U}_i(\bm{\xi}_i;\bm{\theta}_i)\right>=\frac{3}{2}\Omega_{0}
(1-\bar{\alpha})  \frac{(1+z_i)^{2} d^{2}(0;z) \Delta z_i}{Y(z_i)}. \label{eq:UMii}
\end{equation}
\par
Substituting equation (\ref{eq:UMii}) into equation (\ref{eq:UUU}) and 
using equations (\ref{eq:chijfunc}) and (\ref{eq:chifunc}),  we
have the average magnification matrix 
as follows:
\begin{equation}
\left<A_{\mathrm{S},N_\mathrm{L}}\right>
=\left[1+
\sum_{k=1}^{N_\mathrm{L}} 
\left(-\right)^k
\sum_{i_1=1}^{i_0}\Delta \tau_{i_1,i_0}\sum_{i_2=1}^{i_1}\Delta \tau_{i_2,i_1}
\cdots
\sum_{i_k=1}^{i_{k-1}}\Delta \tau_{i_k,i_{k-1}}
\right]I,
\label{eq:As}
\end{equation}
where
\begin{equation}
\Delta \tau_{i,j}=\frac{3}{2}\Omega_0(1-\bar{\alpha})
{(1+z_i)^2d(0;z_i)d(z_i;z_j)\Delta z_i\over d(0;z_j)Y(z_i)},
\label{eq:tau}
\end{equation}
which is the optical depth from the $i$-th lens plane to the $j$-th lens plane 
($i<j$).
\section{Continuous limit}
Keeping the total mass of lenses in the universe up to the redshift
$z_\mathrm{S}$ to be constant, we consider the limit of 
$N_\mathrm{L}\to\infty$. In the case of the infinite number of lenses 
the redshift interval $\Delta z_i$ from the $i$-th lens to the
($i+1$)-th lens plane becomes to be infinitesimal and  then the lens 
plane are distributed continuously up to the redshift $z_\mathrm{S}$ .   
\par
Thus, in this continuous limit, summations with respect to $i$s in 
equation (\ref{eq:As}) become to integrations with respect to $z_i$s, 
respectively, and the average magnification matrix is found to be given 
by 
\begin{equation}
\left<A_\mathrm{S}\right>=
\lim_{N_\mathrm{L}\to\infty}\left<A_{\mathrm{S},N_\mathrm{L}}\right>=B(z_\mathrm{S})I,
\end{equation}
where the function $B(z_\mathrm{S})$ is defined as
\begin{eqnarray}
B(z_\mathrm{S})&\equiv&1+\sum_{k=1}^\infty\left\{-\frac{3}{2}\Omega_0(1-\bar{\alpha})\right\}^k
\int_0^{\zeta_0}d\zeta_1
{(1+\zeta_1)^2d(0;\zeta_1)d(\zeta_1;\zeta_0)\over d(0;\zeta_0)Y(\zeta_1)}\nonumber\\
&\times&\int_0^{\zeta_1}d\zeta_2
{(1+\zeta_2)^2d(0;\zeta_2)d(\zeta_2;\zeta_1)\over d(0;\zeta_1)Y(\zeta_2)}\cdots
\int_0^{\zeta_{k-1}}d\zeta_k
{(1+\zeta_k)^2d(0;\zeta_k)d(\zeta_k;\zeta_{k-1})\over d(0;\zeta_{k-1})Y(\zeta_k)},
\label{eq:BInt}
\end{eqnarray}
and $\zeta_0\equiv z_\mathrm{S}$. 
Equation (\ref{eq:BInt}) can be rewritten in form of the integral equation
\begin{equation}
B(z_\mathrm{S})=1-\frac{3}{2}\Omega_0(1-\bar{\alpha})
\int_0^{z_\mathrm{S}}{(1+z)^2d(0;z)d(z;z_\mathrm{S})\over d(0;z_\mathrm{S})Y(z)}B(z)dz,
\label{eq:BBInt}.
\end{equation}
\par
From equation (\ref{eq:BBInt}) it can be shown that $B(z)$ satisfy the 
differential equation 
\begin{equation}
{d\ \over dz}\left\{(1+z)^2d^2(0;z)Y(z){d\ \over dz}B(z)\right\}+\frac{3}
 {2}\Omega_0(1-\bar{\alpha}){(1+z)^3d^2(0;z)\over Y(z)}B(z)=0,
\label{eq:Beq}
\end{equation}
with the initial conditions 
\begin{equation}
B(z)\Bigr|_{z=0}=1,\qquad {d\ \over dz}B(z)\Bigr|_{z=0}=0.
\label{eq:BIcon}
\end{equation}
The differential equation (\ref{eq:Beq}) can also be obtained by taking 
the continuous limit of equation (\ref{eq:MagBN}).
\par
Since the average magnification factor $\left<\mu(z)\right>$ is given by 
$\left<\mu(z)\right>=B^{-2}(z)$, now we  define a new angular diameter distance 
$\tilde{d}(0;z)$ from  
the observer to a source at $z$ in terms of $d(0;z)$ and $B(z)$ as
\begin{equation}
\tilde{d}(0;z)\equiv \left<\mu(z)\right>^{-1/2}d(0;z)=B(z)d(0;z),
\end{equation}
which is the angular diameter distance magnified with the gravitational 
lens effect.
From equations (\ref{eq:Beq}),(\ref{eq:BIcon}) and (\ref{eq:GD}), it 
follows that $\tilde{d}(0;z)$ satisfies the following differential 
equation 
\begin{equation}
{d\ \over dz}\left\{(1+z)^2Y(z){d\ \over dz}\tilde{d}(0;z)\right\}
+\frac{3}{2}\Omega_0{(1+z)^3\over Y(z)}\tilde{d}(z)=0,
\label{eq:Dneweq}
\end{equation}
and boundary conditions
\begin{equation}
\tilde{d}(0;z)\Bigr|_{z=0}=0,\qquad {d\ \over dz}\tilde{d}(0;z)\Bigr|_{z=0}=1.
\label{eq:Dnewcon}
\end{equation}
\par
Equations (\ref{eq:Dneweq}) and (\ref{eq:Dnewcon}) are the same as 
equation (\ref{eq:GD}) with $\bar{\alpha}=1$. Thus we showed that the newly
defined angular-diameter distance $\tilde{d}(0;z)$ is equivalent to the 
angular diameter distance $d_\mathrm{FL}(0;z)$ in the Friedmann--Lema\^{\i}tre 
universe with the density parameter $\Omega_{0}$ in which all matter 
density is smoothly distributed.
\section{Discussion and Conclusion} 
We have to notice that equations (\ref{eq:BInt}) -- (\ref{eq:Beq}),
(\ref{eq:Dneweq}) and (\ref{eq:Dnewcon}) hold for arbitrary values of 
$\Omega_0$ and of $\bar{\alpha}$.  
In the case of the universe with $\Omega_0 \ll 1$, however, the right 
hand side of equation (\ref{eq:BInt}) can be understood as the expansion 
into power series of $\Omega_0$. The second term  $B_1(z)$ of the 
expansion is given by  
\begin{equation}
B_1(z)=-\frac{3}{2}\Omega_0(1-\bar{\alpha})\int_0^z
{(1+\zeta_1)^2d(0;\zeta_1)d(\zeta_1;z)\over d(0;z)Y(\zeta_1)}d\zeta_1,\label{eq:B1}
\end{equation}
which gives the gravitational magnification effect caused by one 
deflection. It is interesting that $-B_1(z)$ is identical to the optical 
depth $\tau(z)$ introduced by \cite{VO83}. In the case of $\Omega_0\ll1$ 
it is sufficient to take $B_1(z)$ into account in order to obtain 
$\tilde{d}(0;z)$, which is the result discussed by Weinberg (with 
$\bar{\alpha}=0$). 
\par
\begin{figure}[hbt]
\epsscale{.7}
\plotone{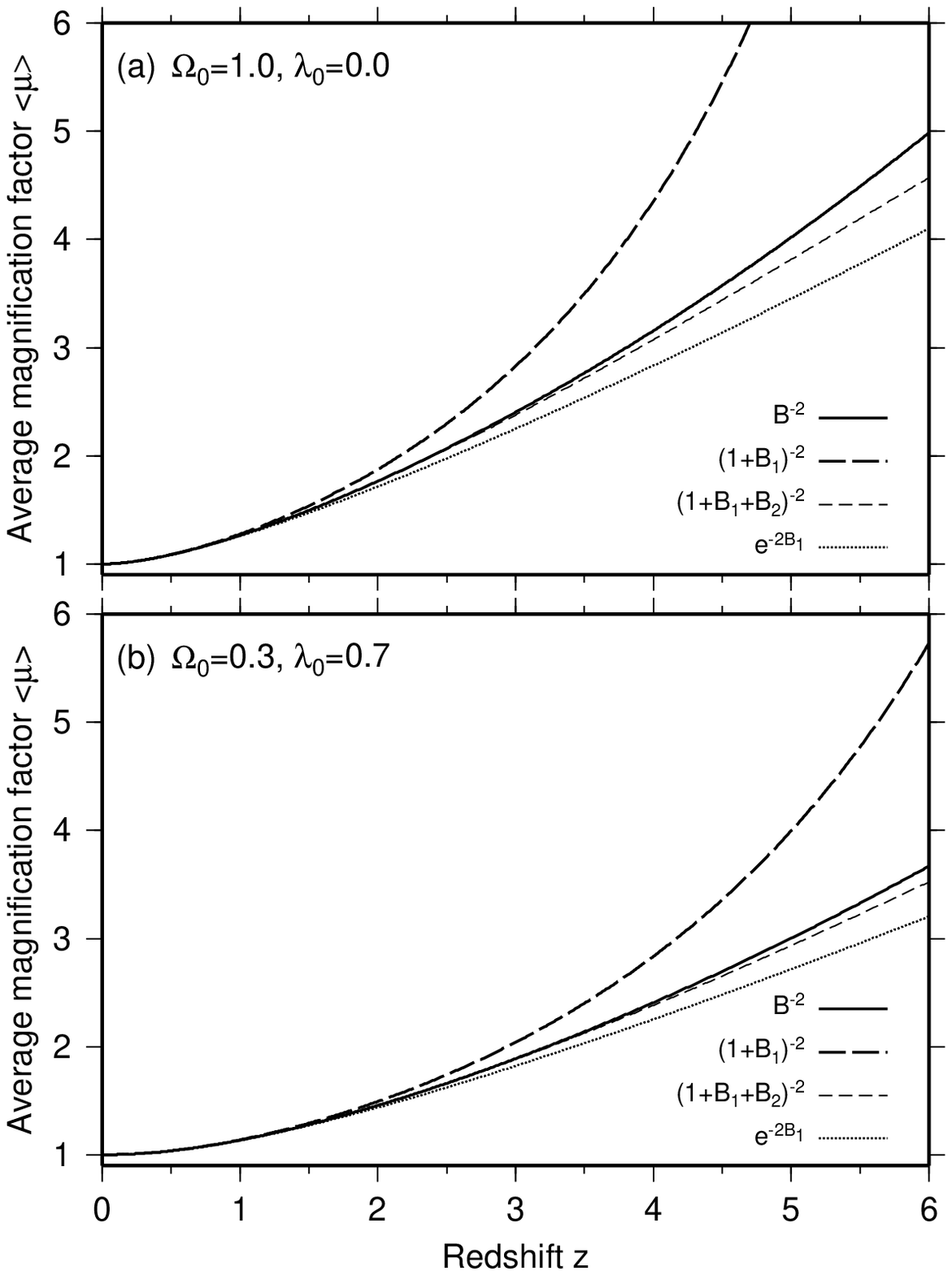}
\caption{Average magnification factor $\left<\mu\right>$: Figure (a) 
 shows the average magnification factor in the model of 
 $\Omega_0=1.0, \lambda_0=0.0$. Figure (b) does in the model of
 $\Omega_0=0.3, \lambda_0=0.7$. The thick solid line, thick long-dashed 
 line, dashed line and dotted line line show
 $B^{-2}(z)$, $(1+B_1)^{-2}=(1-\tau)^{-2}$,
 $(1+B_1+B_2)^{-2}$ and $e^{-2B_1(z)}=e^{2\tau}$, 
 respectively.
\label{fig:Bs}
}
\end{figure}

In a general case of the universe with arbitrary $\Omega_0$ and 
$\bar{\alpha}$, we have to consider $B(z)$ itself which includes 
gravitational magnification effects caused by multiple deflections. The 
third term $B_2(z)$ in the right hand side of equation (\ref{eq:BInt}), 
for example, is written by 
\begin{equation}
B_2(z)=\left[-\frac{3}{2}\Omega_0(1-\bar{\alpha})\right]^2\int_0^z
{(1+\zeta_1)^2d(0;\zeta_1)d(\zeta_1;z)\over d(0;z)Y(\zeta_1)}d\zeta_1\int_0^{\zeta_1}
{(1+\zeta_2)^2d(0;\zeta_2)d(\zeta_2;\zeta_1)\over d(0;\zeta_1)Y(\zeta_2)}d\zeta_2,
\label{eq:B2}
\end{equation}
which is not equal to $\bigl[B_1(z)\bigr]^2/2$. In the same manner the 
$(n+1)$-th term $B_n(z)$ is found not to be equal 
$[B_1(z)]^{n}/n!$. This comes from the fact the total 
magnification by the multiple deflections can not be given by a product 
of the magnifications by individual deflectors. We have to notice  that 
our average magnification factor $\left<\mu\right>$ coincides neither 
with $[1-\tau(z)]^{-2}$ given by \cite{Young} nor with $e^{2\tau(z)}$ 
obtained by \cite{Pei93}.  
These differences, however, are not significant in the range with $z\la1$, but 
become not to be negligible in the range with $z>1$ even in the case of 
$\Omega_0<1$ (see Fig. \ref{fig:Bs}).
\par
Since our universe is locally inhomogeneous, on-average homogeneous, it 
is needed to know how a light ray propagate in the universe. 
Unfortunately we have no such cosmological model derived from the 
Einstein equation, then we have to investigate which working model to be 
plausible is reasonable and useful to discuss the problem. In this 
point of view, our conclusion that the gravitationally magnified angular 
diameter distance $\tilde{d}(0;z)$ reduces to $d_\mathrm{FL}(0;z)$ in the 
continuous limit is the important result which guarantees the fact that 
the hypothetical clumpy universe taken the gravitational lens effects 
into account is the reasonable working  model to study the light ray 
propagation in the inhomogeneous universe\footnote{
\cite{SEF92} have discussed in their book the continuous limit of the 
multi-plane lens equation with the negative surface mass densities and showed the 
Dyer-Roeder angular diameter distance can be derived from their model.}. 
\par
\acknowledgments
This research was partially supported by the Grant-in-Aid for Scientific Research on 
Priority Areas (13135218) of the Ministry of Education, Science, Sports 
and Culture of Japan.
\appendix
\section{Background Universe Model and Angular Diameter Distance}
The metric of the Friedmann--Lema\^{\i}tre universe (the FL universe) is given by
\begin{equation}
ds^2=c^2dT^2-a^2(T)\left({dr^2\over1-kr^2}+r^2d\theta^2+r^2\sin^2\theta 
d\phi^2\right),
\end{equation}
where $T$ is the cosmological time, and $a(T)$ is the expansion factor
which has the dimension of distance. 
The Einstein equation in this geometry yields the following relation
\begin{equation}
\left({\dot{a}\over a}\right)^2={8\pi 
G\over3}\bar{\rho}(z)+{\Lambda c^2\over3}-{kc^2\over a^2},
\label{eq:A3}
\end{equation}
where $\bar{\rho}(z)$ and $\Lambda$ are the mean density of the 
universe at redshift $z$ and the cosmological constant, respectively, 
and the dot denotes the derivative with $T$.
Since $\bar{\rho}(z)$ and $a(T)$ are given in terms of their 
present values $\bar{\rho}_0$ and $a_0$ by 
$\bar{\rho}(z)=\bar{\rho}_0(1+z)^3$ and $a(T)=a_0/(1+z)$,  
respectively, equation (\ref{eq:A3}) is rewritten as
\begin{equation}
\left({\dot{a}\over a}\right)^2
 =H_0^2\left[\Omega_0(1+z)^3+\lambda_0-K(1+z)^2\right] 
\equiv H_0^2Y^2(z),
\label{eq:expand}
\end{equation}
where  $\Omega_0=8\pi G\bar{\rho}_0/3H_0^2$,
$\lambda_0=\Lambda c^2/3H_0^2, K=kc^2/H_0^2a_0^2(=\Omega_0+\lambda_0-1)$ 
and $H_0$ is the present Hubble constant.
\par
It follows from equation (\ref{eq:expand}) that
the relation between $T$ and $z$ is expressed as
\begin{equation}
-cdT={c\over H_0}{dz\over (1+z)Y(z)}.\label{eq:Tredshift}
\end{equation}
Furthermore the relation between $z$ and an affine 
parameter $v$ along a light ray is derived from equation 
(\ref{eq:Tredshift}) and from the geodesic equation for the light ray as follows:
\begin{equation}
 dv={dz\over(1+z)^2Y(z)}.
 \end{equation}
The dimensionless angular diameter distance $d(z_a;z_b)$  from a lens at 
redshift $z_a$ to another at $z_b$ of the clumpy
universe in which a mass fraction $\bar{\alpha}$ of the mean matter 
density is smoothly distributed satisfies the following equation 
\begin{equation}
(1+z_b)^2Y(z_b)
{d\ \over dz_b}\left\{
(1+z_b)^2Y(z_b){d\ \over dz_b}d(z_a;z_b)\right\}
+\frac{3}{2}\Omega_0\bar{\alpha} (1+z_b)^5
d(z_a;z_b)=0,\label{eq:GD}
\end{equation} 
and the following initial condition
\begin{equation}
d(z_a;z_b)\Bigr|_{z_b=z_a}=0,\qquad {d \over dz_b}d(z_a;z_b)\Bigr\vert_{z_b=z_a}
={1\over (1+z_a)Y(z_a)},
\end{equation}
\citep{DR73}.
The second condition is the Hubble law at redshift $z_a$ \citep{SEF92}.
\par
\cite{SEF92} define the $\chi$-function in order to express a time 
delay function by
\begin{equation}
\chi_{ab}={(1+z_a)d(0;z_a)d(0;z_b)\over 
d(z_a;z_b)}={1\over\chi_a-\chi_b}.\label{eq:chijfunc} 
\end{equation}
The relation between the $\chi$-function and redshift 
$z$ is also rewritten as
\begin{equation}
\chi_a=\chi(z_a)=\int_{z_a}^\infty {dz\over (1+z)^2Y(z)d^2(0;z)},\label{eq:chifunc}
\end{equation}
\citep*[see also][]{SSE94}.
\section{Proof of equation (\ref{eq:UUU})}
In this appendix, we prove that the equation (\ref{eq:UUU}) holds. 
In equation (\ref{eq:Uip}), the matrix $\tilde{U}_i$ is a function of 
$\bm{\theta}_i-\bm{\xi}_i-\tilde{\bm{\theta}}_i$. Then, by transforming 
variables $\bm{\xi}_i$ to 
$\bm{\phi}_i\equiv\bm{\xi}_i+\tilde{\bm{\theta}}_i-\bm{\theta}_i$ and putting 
$V_i(\bm{\phi}_i)=\tilde{U}_i(\bm{\xi}_i;\bm{\theta}_i)$, we have 
\begin{eqnarray}
&&\iint_{\cal D}d^2\bm{\xi}_{i_1}\cdots\iint_{\cal D}d^2\bm{\xi}_{i_k}
\tilde{U}_{i_1}(\bm{\xi}_{i_1};\bm{\theta}_{i_1})
\tilde{U}_{i_2}(\bm{\xi}_{i_2};\bm{\theta}_{i_2})\cdots 
\tilde{U}_{i_k}(\bm{\xi}_{i_k};\bm{\theta}_{i_k})
\nonumber\\
&=&\iint_{\cal D}d^2\bm{\phi}_{i_1}\cdots\iint_{\cal D}d^2\bm{\phi}_{i_k}
\Biggl|
{\partial(\bm{\xi}_{i_1}\cdots\bm{\xi}_{i_k})
\over
\partial(\bm{\phi}_{i_1}\cdots\bm{\phi}_{i_k})}
\Biggr|
V_{i_1}(\bm{\phi}_{i_1})\cdots 
V_{i_k}(\bm{\phi}_{i_k}),\label{eq:avUUU}
\end{eqnarray}
where is
$\partial(\bm{\xi}_{i_1}\cdots\bm{\xi}_{i_k})/
\partial(\bm{\phi}_{i_1}\cdots\bm{\phi}_{i_k})$
is the Jacobian matrix.
We define a $2\times2$ sub-Jacobian matrix $J_k^i$ as
\begin{displaymath}
J_k^i={\partial \bm{\phi}_i\over\partial\bm{\xi}_k},
\end{displaymath}
then the inverse Jacobian matrix can be written as follows:
\begin{equation}
{\partial(\bm{\phi}_{i_1}\cdots\bm{\phi}_{i_k})\over
\partial(\bm{\xi}_{i_1}\cdots\bm{\xi}_{i_k})}
=\left(
\begin{array}{ccc}
J_{i_1}^{i_1}&\ldots&J_{i_1}^{i_k}\\
\vdots&\ddots&\vdots\\
J_{i_k}^{i_1}&\ldots&J_{i_k}^{i_k}\\
\end{array}
\right).\label{eq:Jacb}
\end{equation}
From equation (\ref{eq:mlptheta}) it can be shown that $\bm{\theta}_i$ 
does not depend on $\bm{\xi}_k$ for $k\ge i$ and then that 
$J_{i_1}^{i_1}=J_{i_2}^{i_2}=\cdots=J_{i_k}^{i_k}=I$ and 
$J_{i_k}^{i_l}=O$ for $i_l>i_k$. Thus we find the inverse Jacobian 
matrix has a form given by 
\begin{equation}
{\partial(\bm{\phi}_{i_1}\cdots\bm{\phi}_{i_k})\over
\partial(\bm{\xi}_{i_1}\cdots\bm{\xi}_{i_k})}
=\left(
\begin{array}{cccc}
I&J_{i_1}^{i_2}&\ldots&J_{i_1}^{i_k}\\
O&I&\ddots&\vdots\\
\vdots&\ddots&\ddots&J_{i_{k-1}}^{i_k}\\
O&\ldots&O&I\\
\end{array}
\right).
\end{equation}
This matrix is a upper triangle matrix of which diagonal component is 1 
and then the determinant of the matrix is unity. And therefore the
determinant of the inverse matrix which is the Jacobian of mapping 
$\bm{\xi}_i\to\bm{\phi}_i=\bm{\xi}_i-\bm{\theta}_i-\bm{\eta}_i$ is unity, 
too. Thus the right hand side of equation (\ref{eq:avUUU}) can be 
rewritten as
\begin{eqnarray}
&&\iint_{\cal D}d^2\bm{\phi}_{i_1}\cdots\iint_{\cal D} d^2\bm{\phi}_{i_k}
\Biggl|
{\partial(
\bm{\xi}_{i_1}\cdots\bm{\xi}_{i_k}
)\over\partial(\bm{\phi}_{i_1}\cdots\bm{\phi}_{i_k})}
\Biggr|
V_{i_1}(\bm{\phi}_{i_1})\cdots V_{i_k}(\bm{\phi}_{i_k})\nonumber\\
&&\hspace{20mm}=\left[\iint_{\cal D}V_{i_1}(\bm{\phi}_{i_1})d^2\bm{\phi}_{i_1}\right]\cdots
\left[\iint_{\cal D}V_{i_k}(\bm{\phi}_{i_k})d^2\bm{\phi}_{i_k}\right].
\label{eq:Vprod}
\end{eqnarray}
From (\ref{eq:avUUU}) and (\ref{eq:Vprod}) we have equation (\ref{eq:UUU}).


\begin{thebibliography}{}
\bibitem[Blandford \& Narayan(1986) Blandford \& Narayan]{BL86} Blandford, D.R. \& Narayan,
					 R. 1986, \apj, 310,568
\bibitem[Dyer \& Roeder(1973) Dyer \& Roeder]{DR73} Dyer, C.C. \& Roeder, R.C. 1973,
					 \apjl, 180, L31
\bibitem[Ehlers \& Schneider(1986) Ehlers \& Schneider]{ES86} Ehlers, J. \& Schneider, P.
					 1986, \aap, 168, 57
\bibitem[Isaacson \& Canizares(1989)]{Isaac89} Isaacson, J.A. \& 
Canizares, C.R. 1989,\apj,336,544
\bibitem[Kovner(1987) Kovner]{Kovn87} Kovner,I. 1987,\apj,316,52
\bibitem[Lee et al.(1997) Lee, Babul, Kofman \& Kaiser]{Lee97} Lee, M.H, 
Babul, A., Kofman, L. \& Kaiser, N. 1997,\apj,489,522
\bibitem[Marchandon \& Nottale(1991)]{Marchan} Marchandon, S. \& Nottale, 
L. 1991,\aap,251,393
\bibitem[Omote \& Yoshida(1990)]{OY90}Omote, M. \& Yoshida, H. 1990,
					 \apj, 361, 27
\bibitem[Peacock(1986) Peacock]{Pea86} Peacock, J.A. 1986, \mnras, 233,
					 113
\bibitem[Rauch(1990)]{Rauch90} Rauch, K.P. 1991,\apj,374,83
\bibitem[Refsdal(1970)]{Refs70} Refsdal, S. 1970,\apj,159,357
\bibitem[Pei(1993)]{Pei93} Pei,Y.,C. 1993, \apj, 403, 7
\bibitem[Perlmutter et al.(1999)]{SNIa} Perlmutter,S. et al. 1999,\apj,517,565
\bibitem[Petters et al.(2001) Petters, Levine \& Wambsganss]{PLW01}
					 Petters, A.O., Levine, H. \&
					 Wambsganss, J. 2001,
					 Singularity Theory and
					 Gravitational Lensing
					 (Birkh\"{a}user) 
\bibitem[Premadi et al.(2001) Premadi, Martel, Matzner \& 
Futamase]{PMMF01} Premadi, P., Martel, H., Matzner, R. \& Futamase, T. 
2001,\apjs,135,7
\bibitem[Seitz \& Schneider(1994)]{SS94} Seitz, S. \& Schneider,P.
					 1994,\aap,287,349
\bibitem[Schneider(1993)]{Schneider93} Schneider, P. 1993,\aap,278.1
\bibitem[Schneider et al.(1992) Schneider, Ehlers \& Falco]{SEF92}
					 Schneider, P., Ehlers, J. \&
					 Falco, E.E. 1992,
					 Gravitational Lenses
					 (Springer-Verlag)
\bibitem[Schneider \& Weiss(1988a)]{SW88a} Schneider, P. \& Weiss, A. 
1988a, \apj,327,440
\bibitem[Schneider \& Weiss(1988b)]{SW88b} Schneider, P. \& Weiss, A. 
1988b, \apj,330,1
\bibitem[Seitz et al.(1994) Seitz, Schneider \& Ehlers]{SSE94} Seitz,
					 S., Schneider, P., \& Ehlers,
					 J. 1994, Class.Quant.Grav.,
					 11, 2345
\bibitem[Spergel et al.(2003)]{WMAP} Spergel, D.N. et al. 2003,preprint (astro-ph/0302209)
\bibitem[Vietri \& Ostriker(1983)]{VO83} Vietri, M. \& Ostriker, J.P.
					 1983, \apj, 267, 488
\bibitem[Watanabe \& Tomita(1990)]{WT90} Watanabe, K. \& Tomita, K. 1990,\apj,355,1
\bibitem[Weinberg(1976) Weinberg]{Wein76} Weinberg,S
					 1976,\apj,208,L1. 
\bibitem[Wu(1990)]{Wu90} Wu, X. 1990,\aap,232,3
\bibitem[Yoshida \& Omote(1992)]{YO92} Yoshida, H. \& Omote, M. 1992,\apjl,388,L1
\bibitem[Young(1981)]{Young} Young, P. 1981,\apj,244,756
\end{thebibliography}
\end{document}